\documentclass[pra,aps,showpacs,twocolumn]{revtex4}
\usepackage{amssymb}
\usepackage{amsmath}
\usepackage{mathtools}
\usepackage[dvipdfm]{hyperref}
\usepackage{hyperref}

\allowdisplaybreaks
\begin{document}

\title{Coherence measures with respect to general quantum measurements
}
\author{Jianwei Xu\textsuperscript{1}}
\email{xxuianwei@nwafu.edu.cn}
\author{Lian-He Shao\textsuperscript{2}}
\author{Shao-Ming Fei\textsuperscript{3,4}}
\email{feishm@cnu.edu.cn}
\affiliation{\textsuperscript{1}College of Science, Northwest A\&F University, Yangling, Shaanxi 712100,
China}
\affiliation{\textsuperscript{2}School of Computer Science, Xi'an Polytechnic University, Xi'an, 710048, China}
\affiliation{\textsuperscript{3}School of Mathematical Sciences, Capital Normal University, Beijing 100048, China}
\affiliation{\textsuperscript{4}Max-Planck-Institute for Mathematics in the Sciences, Leipzig 04103, Germany}

%\date{\today }

\begin{abstract}

Quantum coherence with respect to orthonormal bases has been
studied extensively in the past few years. Recently, Bischof, et al. [Phys. Rev. Lett. 123,
110402 (2019)] generalized it to the case of general positive operator-valued measure (POVM)
measurements. Such POVM-based coherence, including the block coherence
as special cases, have significant operational interpretations in quantifying
the advantage of quantum states in quantum information processing. In this work
we first establish an alternative framework for quantifying
the block coherence and provide several block coherence measures.
We then present several coherence measures with respect to
POVM measurements, and prove a conjecture on the $l_{1}$-norm related POVM coherence measure.
\end{abstract}

\pacs{03.65.Ud, 03.67.Mn, 03.65.Aa}
\maketitle

\section{Introduction}

Quantum coherence is a characteristic feature of quantum mechanics, with wide
applications in superconductivity, quantum thermodynamics and
biological processes. From a resource-theoretic perspective
the quantification of quantum coherence has attracted
much attention and various kinds of coherence measures
have been proposed \cite{BCP-2014-PRL,XY,AW,CN,LZ1,LZ2,BC,KB1,Yu-2017-PRA,LUO8,LUO9,LUO10,Wu-2018-PRA,Xu-2020-CPB,XNZ}.
Let $\rho$ be a density operator in $d$-dimensional complex Hilbert space $H$.
Under a fixed orthonormal basis $\{|i\rangle \}_{i=1}^{d}$ of $H$,
the state $\rho $ is called incoherent if $\langle i|\rho |j\rangle =0$ for any $i\neq j$ \cite{BCP-2014-PRL}. Otherwise $\rho$ is called coherent.
The coherence theory has achieved fruitful results in the past few years (for recent reviews
see e.g. \cite{Plenio-2016-RMP,Fan-2018-PhysicsReports}).

From another perspective, the orthonormal basis $\{|i\rangle \}_{i=1}^{d}$
corresponds to a rank-1 projective measurement (von Neumann measurement) $%
\{|i\rangle \langle i|\}_{i=1}^{d}$, and $\langle i|\rho |j\rangle =0$ is
equivalent to $|i\rangle \langle i|\rho |j\rangle \langle j|=0$. This
observation leads one to view the coherence with respect to the orthonormal
basis $\{|i\rangle \}_{i=1}^{d}$ as the coherence with respect to the rank-1
projective measurement $\{|i\rangle \langle i|\}_{i=1}^{d}$. Along this idea,
the concept of coherence can be generalized to the cases of general measurements.
Recently, Bischof, et al. \cite{Brub-2019-PRL} have generalized the concept of
coherence to the case of general quantum measurements, i.e.,
positive operator-valued measures (POVMs).
One motivation of this
generalization is due to the fact that POVMs may be more advantageous compared to
von Neumann measurement in many applications \cite{Acin-2017-PRL}. There are many important problems, such as the optimal way to distinguish a set of quantum states, involve POVM, rather than projective measurement.
Moreover, the notion of coherence with respect to a general measurement
can be embedded in a consistent resource theory, and such POVM-based coherence measures
have interesting operational interpretations which quantify the advantage of quantum states
in a quantum information protocol \cite{Brub-2019-arxiv}. Refs. \cite{Brub-2019-PRL,Brub-2019-arxiv} provided a way of generalizing coherence theory not in an orthonormal basis, but with a generic POVM. This effort has been started in \cite{Plenio-PRL-2016,Plenio-PRL-2017}.

After establishing a framework for quantifying the POVM coherence
\cite{Brub-2019-PRL,Brub-2019-arxiv}, Bischof, et
al. developed \cite{Brub-2019-PRL,Brub-2019-arxiv} a
scheme by employing the Naimark extension to embed the POVM coherence into
the block coherence proposed in \cite{Aberg-2006-arxiv} in a lager Hilbert space.
The Naimark extension \cite{Peres-2006-book,Decker-2005-JMP} states that any POVM can
be extended to a projective measurement in a larger Hilbert space. The block
coherence was defined with respect to projective measurements, not
necessarily rank-1. With this scheme, the relative entropy of POVM coherence
$C_{\text{rel}}$, the robustness POVM coherence $C_{\text{rot}}$ were proposed.
Recently, the structures of different incoherent operations for POVM coherence were investigated \cite{Dey-2018-arxiv}.
For simplicity, we call the coherence theory with respect to fixed
orthonormal bases the standard coherence theory. As the generalizations of the
standard coherence, both the block coherence and the POVM coherence
reduce to the standard coherence in the case of the von Neumann
measurement.

In the present work, we establish an alternative framework for quantifying
the block coherence and provide several block coherence measures.
We then present several POVM coherence measures. Meanwhile, we also prove a conjecture
raised recently.

\section{Alternative framework for quantifying block coherence}

\subsection{Block incoherent states and block incoherent channels}

The block coherence theory was introduced in \cite{Aberg-2006-arxiv}.
We adopt the framework proposed in \cite{Brub-2019-arxiv} for quantifying the block coherence.
Consider a quantum system $A$ associated with an $m$-dimensional
complex Hilbert space $H$. One has partition $H=\oplus _{i=1}^{n}\pi _{i}$ into
orthogonal subspaces $\pi _{i}$ of dimension dim$\pi _{i}=m_{i}$,
$\sum_{i=1}^{n}m_{i}=m$. Correspondingly, one gets a projective
measurement $P=\{P_{i}\}_{i=1}^{n}$, with each projector satisfying
$P_{i}(H)=\pi _{i}$. A state $\rho $ on $H$ is called block incoherent (BI) with
respect to $P$ if
\begin{eqnarray}
P_{i}\rho P_{j}=0,~~~\forall i\neq j,  \label{eq1}
\end{eqnarray}
or
\begin{eqnarray}
\rho =\sum_{i=1}^{n}P_{i}\rho P_{i}.  \label{eq2}
\end{eqnarray}
We denote the set of all quantum states in $H$ by $\mathcal{S}(H)$, and
the set of all block incoherent quantum states by $\mathcal{I}_{\text{B}}(H)$.
It is easy to check that
\begin{eqnarray}
\mathcal{I}_{\text{B}}(H)=\{\sum_{i=1}^{n}P_{i}\rho P_{i}|\rho \in \mathcal{%
S}(H)\}.   \label{eq3}
\end{eqnarray}

A quantum channel is a completely positive and trace preserving (CPTP) linear map of quantum states \cite{Nielsen-2000-book}. A quantum
channel $\phi $ is often expressed by the Kraus operators $\{K_{l}\}_{l}$
satisfying $\sum_{l}K_{l}^{\dagger }K_{l}=I_{m}$, where $I_{m}$ is the identity operator on $H$ and $\dag $ stands for the adjoint. A quantum channel $\phi $ is called block incoherent if it admits an
expression of Kraus operators $\phi =\{K_{l}\}_{l}$ such that
\begin{eqnarray}
P_{i}K_{l}\rho K_{l}^{\dagger }P_{j}=0,~~~\forall l,~\forall i\neq j \label{eq4}
\end{eqnarray}
for any $\rho \in\mathcal{I}_{\text{B}}(H)$.
Such an expression $\phi =\{K_{l}\}_{l}$ is called a block
incoherent decomposition of $\phi$.
We denote the set of all quantum channels on $H$ by $\mathcal{C}(H)$, and
the set of all block incoherent quantum channels by $\mathcal{C}_{\text{BI}}(H)$.

The concept of block coherence can be properly extended to the multipartite
systems via the tensor product of the Hilbert spaces of the subsystems, similar to
the case of standard coherence theory \cite{Plenio-2016-RMP}.
For bipartite systems, let $A^{\prime }$ be another quantum system associating with the
$m^{\prime }$-dimensional complex Hilbert space $H^{\prime }$.
Partitioning $H^{\prime }=\oplus _{i=1}^{n^{\prime }}\pi
_{i}^{\prime }$ into orthogonal subspaces $\pi _{i}^{\prime }$ of
dimension dim$\pi _{i}^{\prime }=m_{i}^{\prime }$,
$m^{\prime }=\sum_{i=1}^{n^{\prime }}m_{i}^{\prime }$, one gets a
projective measurement $P^{\prime }=\{P_{i}^{\prime }\}_{i=1}^{n^{\prime }}$
with each projector $P_{i}^{\prime }$ satisfying $P_{i}^{\prime }(H^{\prime })=\pi
_{i}^{\prime }$. Correspondingly one has concepts such as $%
\mathcal{S}(H^{\prime }),$ $\mathcal{I}_{\text{B}}(H^{\prime }),$ $\mathcal{C%
}(H^{\prime })$ and $\mathcal{C}_{\text{BI}}(H^{\prime }).$
For the composite Hilbert space $H^{AA^{\prime }}=H^{A}\otimes H^{A^{\prime }}$ associating to
the bipartite system $AA^{\prime }$, we have the projective measurement $%
P\otimes P^{\prime }=\{P_{i}\otimes P_{i^{\prime }}^{\prime }\}_{ii^{\prime
}}$. A state $\rho ^{AA^{\prime }}$ on $H^{AA^{\prime }}$ is called block
incoherent with respect to the projective measurement $P\otimes P^{\prime }$
if
\begin{eqnarray}
(P_{i}\otimes P_{i^{\prime }}^{\prime })\rho ^{AA^{\prime }}(P_{j}\otimes
P_{j^{\prime }}^{\prime })=0,~~~\forall (i,i^{\prime })\neq (j,j^{\prime }),  \label{eq5}
\end{eqnarray}
where $(i,i^{\prime })\neq (j,j^{\prime })$ means that $i\neq j$ or $%
i^{\prime }\neq j^{\prime }.$

We denote the set of all states on $%
H^{AA^{\prime }}$ by $\mathcal{S}(H^{AA^{\prime }})$ and the set of
all channels on $\mathcal{S}(H^{AA^{\prime }})$ by $\mathcal{C}%
(H^{AA^{\prime }}).$ A quantum channel $\phi ^{AA^{\prime }}$ on $\mathcal{C}%
(H^{AA^{\prime }})$ is called a block incoherent if it admits an expression
of Kraus operators $\phi ^{AA^{\prime }}=\{K_{l}^{AA^{\prime }}\}_{l}$ such
that
\begin{eqnarray}
(P_{i}\otimes P_{i^{\prime }}^{\prime })K_{l}^{AA^{\prime }}\rho
^{AA^{\prime }}(K_{l}^{AA^{\prime }})^{\dagger }(P_{j}\otimes P_{j^{\prime
}}^{\prime })=0 \label{eq6}
\end{eqnarray}
for all $l$ and $(i,i^{\prime })\neq (j,j^{\prime })$.
We denote the set of all block incoherent channels on $\mathcal{C}%
(H^{AA^{\prime }})$ by $\mathcal{C}_{\text{BI}}(H^{AA^{\prime }})$ and call
such an expression $\phi ^{AA^{\prime
}}=\{K_{l}^{AA^{\prime }}\}_{l}$ a block incoherent decomposition of $\phi ^{AA^{\prime}}$.

\subsection{An alternative framework for quantifying the block coherence}

A framework for quantifying the block coherence has been established
in \cite{Brub-2019-arxiv}: any valid
block coherence measure $C(\rho ;P)$ with respect to the projective measurement $P$ should satisfy the conditions (B1-B4)
below.

\textbf{(B1)} Faithfulness: $C(\rho ;P)\geq 0$ with equality if $\rho \in \mathcal{I}%
_{\text{B}}(H).$

\textbf{(B2)} Monotonicity: $C(\phi _{\text{BI}}(\rho );P)\leq C(\rho ;P)$ for any $%
\phi _{\text{BI}}\in \mathcal{C}_{\text{BI}}(H).$

\textbf{(B3)} Strong monotonicity: $\sum_{l}p_{l}C(\rho _{l};P)\leq C(\rho ;P)$ for
any block incoherent decomposition
$\phi _{\text{BI}}=\{K_{l}\}_{l}\in \mathcal{C}_{\text{BI}}(H)$ of
$\phi _{\text{BI}}$, $p_{l}=\text{tr}(K_{l}\rho K_{l}^{\dagger })$,
$\rho _{l}=K_{l}\rho K_{l}^{\dagger}/p_{l}$.

\textbf{(B4)} Convexity: $C(\sum_{j}p_{j}\rho _{j};P)\leq \sum_{j}p_{j}C(\rho _{j};P)$
for any states $\{\rho _{j}\}_{j}$ and any probability distribution $%
\{p_{j}\}_{j}.$

This framework coincides with the one in the standard coherence theory
\cite{BCP-2014-PRL} if all $\{P_{i}\}_{i=1}^{n}$ are
rank-1.  Note that (B3) and (B4) together imply (B2).

The framework of the standard coherence theory \cite{BCP-2014-PRL} had been modified by adding an additivity condition in \cite{Tong-2016-PRA}. For the block coherence theory, we add the following
condition:

\textbf{(B5)} Block additivity:
\begin{eqnarray}
C(p_{1}\rho _{1}\oplus p_{2}\rho _{2};P)=p_{1}C(\rho _{1};P)+p_{2}C(\rho
_{2};P),  \label{eq7}
\end{eqnarray}
where $p_{1}>0,$ $p_{2}>0,$ $p_{1}+p_{2}=1,$ $\rho _{1},\rho _{2}\in
\mathcal{S}(H)$, and for any partition $P=\{P_{k_{1}}\}_{k_{1}}\cup
\{P_{k_{2}}\}_{k_{2}}$ such that $\{k_{1}\}_{k_{1}}\cup
\{k_{2}\}_{k_{2}}=\{k\}_{k=1}^{n}$, $\{k_{1}\}_{k_{1}}\cap
\{k_{2}\}_{k_{2}}=\varnothing $ and $\rho _{1}P_{k_{2}}=\rho _{2}P_{k_{1}}=0$
for any $k_{1}$ and $k_{2}.$

With condition (B5), we have the following theorem, which establishes
an alternative framework for quantifying the block coherence.

\textbf{Theorem 1.} The framework given by conditions (B1) to (B4) is equivalent to the one given by the conditions (B1), (B2) and (B5).

{\sf [Proof]}
We first prove that conditions (B1) to (B4) imply (B1), (B2) and (B5).
Suppose that (B1) to (B4) are fulfilled. For the state $p_{1}\rho _{1}\oplus p_{2}\rho _{2}$ as given
in (B5), we construct the BI channel $\phi _{\text{BI}}=\{K_{1},K_{2}\}$ with $%
K_{1}=\sum_{k_{1}}P_{k_{1}}$, $K_{2}=\sum_{k_{2}}P_{k_{2}}.$ We have $%
K_{1}(p_{1}\rho _{1}\oplus p_{2}\rho _{2})K_{1}^{\dagger }=p_{1}\rho _{1}$
and $K_{2}(p_{1}\rho _{1}\oplus p_{2}\rho _{2})K_{2}^{\dagger }=p_{2}\rho
_{2}$. Then from (B3) we get
\begin{eqnarray}
C(p_{1}\rho _{1}\oplus p_{2}\rho _{2};P)\geq p_{1}C(\rho _{1};P)+p_{2}C(\rho
_{2};P).   \label{eqA1}
\end{eqnarray}
On the other hand, since $p_{1}\rho _{1}\oplus p_{2}\rho _{2}=p_{1}\rho
_{1}+p_{2}\rho _{2},$ from (B4) we get
\begin{eqnarray}
C(p_{1}\rho _{1}\oplus p_{2}\rho _{2};P)\leq p_{1}C(\rho _{1};P)+p_{2}C(\rho
_{2};P).  \label{eqA2}
\end{eqnarray}
Combining (\ref{eqA1}) and (\ref{eqA2}) we get the condition (B5).

Next we prove that (B1), (B2) and (B5) imply (B1) to (B4).
Suppose conditions (B1), (B2) and (B5) are satisfied.
Let $\{K_{l}\}_{l=1}^{n^{\prime }}\in\mathcal{C}_{\text{BI}}(H)$ be a BI
decomposition associated to the system $A$.
Consider the bipartite system $AA^{\prime }$ with the
aforementioned notation and $\rho \in \mathcal{S}(H)$. Let the state
$\rho ^{AA^{\prime }}=\rho \otimes |1\rangle \langle 1|$ undergo a BI channel such that
\begin{eqnarray}
\phi _{BI}^{AA^{\prime }}(\rho ^{AA^{\prime }})
&=&\sum_{l}(K_{l}\otimes
U_{l})(\rho \otimes |1\rangle \langle 1|)(K_{l}^{\dagger }\otimes
U_{l}^{\dagger })  \nonumber \\
&=&\sum_{l}K_{l}\rho K_{l}^{\dagger }\otimes |l\rangle \langle l|,  \label{eqA4}
\end{eqnarray}
where
\begin{eqnarray}
U_{l}=\sum_{k=1}^{n^{\prime }}|k+l-1\rangle \langle k|  \label{eqA5} \nonumber
\end{eqnarray}
are the unitary operators on $A^{\prime }$.
From (B5),  (\ref{eqA4}) gives rise to
\begin{eqnarray}
C(\sum_{l}K_{l}\rho K_{l}^{\dagger }\otimes |l\rangle \langle l|;P\otimes
P^{\prime })=\sum_{l}p_{l}C(\rho _{l};P),  \label{eqA6}
\end{eqnarray}
where $P$ and $P^{\prime}$ are rank-1 projective measurements,
$p_{l}=\text{tr}(K_{l}\rho K_{l}^{\dagger })$, $\rho _{l}=K_{l}\rho
K_{l}^{\dagger }/p_{l}$, and we have used
\begin{eqnarray}
C(\rho _{l}\otimes |l\rangle \langle l|;P\otimes P^{\prime })=C(\rho
_{l};P).  \label{eqA7}
\end{eqnarray}
According to (B2), (\ref{eqA4}) and (\ref{eqA6}) together imply (B3).

Now consider the following state
\begin{eqnarray}
\rho^{AA^{\prime}}=\sum_{l=1}^{n^{\prime }}p_{l}\rho _{l}\otimes
|l\rangle \langle l|,  \label{eqA8}
\end{eqnarray}
with $\{p_{l}\}_{l=1}^{n^{\prime }}$ a probability distribution and $\{\rho
_{l}\}_{l=1}^{n^{\prime }}\subset \mathcal{S}(H),$ $\{|l\rangle
\}_{l=1}^{n^{\prime }}$ orthonormal basis of $H^{\prime }.$ According to
(B5), we have
\begin{eqnarray}
C(\sum_{l}p_{l}\rho _{l}\otimes |l\rangle \langle l|;P\otimes P^{\prime
})=\sum_{l}p_{l}C(\rho _{l};P).  \label{eqA9}
\end{eqnarray}
Let $\rho ^{AA^{\prime }}$ undergo a BI channel as
\begin{eqnarray}
\phi_{\text{BI}}^{AA^{\prime}}(\rho^{AA^{\prime}})
&=&\sum_{k=1}^{n^{\prime }}(I^{A}\otimes |1\rangle \langle k|)\rho
^{AA^{\prime }}(I^{A}\otimes |k\rangle \langle 1|)    \nonumber \\
&=&\sum_{j}p_{j}\rho _{j}\otimes |1\rangle \langle 1|.  \label{eqA10}
\end{eqnarray}
Similarly, (B2), (B5), (\ref{eqA9}) and (\ref{eqA10}) together imply (B4).
\hfill \rule{1ex}{1ex}

We have provided an alternative framework for block coherence by proving that the conditions (B1) to (B4) are equivalent to the conditions (B1), (B2) and (B5). The similar condition (B5) in the standard coherence
has particular advantages in calculating coherence of block diagonal states \cite{zhaomj}.
The condition (B5) in the block coherence may also simplify
the calculations of the block coherence for certain block diagonal states.

\subsection{Several block coherence measures}

Under the framework of block coherence above, we now provide several block coherence measures.
Denote $P=\{P_{i}\}_{i=1}^{n}$ a projective measurement on the Hilbert space $H$.
The following Propositions 1-5 provide block coherence measures,
see the detailed proofs in Appendix.

\textbf{Proposition 1.}
$l_{1}$ norm of coherence
\begin{eqnarray}
C_{l_{1}}(\rho ,P)=\sum_{i\neq j}||P_{i}\rho P_{j}||_{\text{tr}}   \label{eq8}
\end{eqnarray}
is a block coherence measure, where $||M||_{\text{tr}}=$tr$\sqrt{M^{\dagger
}M}$ denotes the trace norm of the matrix $M$.

\textbf{Proposition 2.} For $\alpha \in (0,1)\cup (1,2],$ coherence based on Tsallis relative entropy
\begin{eqnarray}
C_{T,\alpha }(\rho ,P)=\frac{1}{\alpha -1}\{\sum_{i}\text{tr}[(P_{i}\rho
^{\alpha }P_{i})^{1/\alpha }]-1\}      \label{eq9}
\end{eqnarray}
is a block coherence measure.

In particular, we have

\textbf{Corollary 1.}
\begin{eqnarray}
\lim_{\alpha \rightarrow 1}C_{T,\alpha }(\rho ,P)=(\ln 2)C_{\text{rel}%
}(\rho ,P),   \label{eq10}
\end{eqnarray}
where
\begin{eqnarray}
C_{\text{rel}}(\rho ,P)=\text{tr}(\rho \log _{2}\rho )-\sum_{i}\text{tr}[(P_{i}\rho
P_{i})\log _{2}(P_{i}\rho P_{i})],  \label{eq11}    \nonumber \\
\end{eqnarray}
and $\ln $ is the natural logarithm.

\textbf{Proposition 3.} Modified trace norm of coherence
\begin{eqnarray}
C_{\text{tr}}(\rho ,P)=\min_{\lambda >0,\sigma \in \mathcal{I}%
_{B}(H)}||\rho -\lambda \sigma ||_{\text{tr}}    \label{eq12}
\end{eqnarray}
is a block coherence measure.

\textbf{Proposition 4.} Coherence weight
\begin{eqnarray}
&&C_{\text{w}}(\rho ,P)   \nonumber \\
&=&\min_{\sigma ,\tau }\{s\geq 0|\rho =(1-s)\sigma +s\tau
,\sigma \in \mathcal{I}_{\text{B}}(H),\tau \in \mathcal{S}(H)\}   \nonumber \\
&=&\min_{\sigma }\{s\geq 0|\rho \geq (1-s)\sigma ,\sigma \in \mathcal{I}%
_{\text{B}}(H)\}   \label{eq13}
\end{eqnarray}
is a block coherence measure.

\textbf{Proposition 5.} For $\alpha \in \lbrack \frac{1}{2},1),$ coherence based on sandwiched R\'{e}nyi relative entropy
\begin{eqnarray}
C_{R,\alpha }(\rho ,P)=1-\max_{\sigma \in \mathcal{I}_{\text{B}}(H)}(\{\text{tr}%
[(\rho ^{\frac{1-\alpha }{2\alpha }}\sigma \rho ^{\frac{1-\alpha }{2\alpha }%
})^{\alpha }]\}^{\frac{1}{1-\alpha }})  \label{eq14}   \nonumber \\
\end{eqnarray}
is a block coherence measure.

When $P$ is a rank-1 projective measurement, $C_{l_{1}}(\rho ,P)$ recovers
the standard coherence measure $C_{l_{1}}(\rho )$ proposed in Ref.
\cite{BCP-2014-PRL}, $C_{T,\alpha }(\rho ,P)$
recovers the standard coherence measure proposed in Ref. \cite{Yu-2018-SR,Yu-2017-PRA,Wu-2018-PRA},  $C_{\text{tr}}(\rho
,P)$ recovers the standard coherence measure proposed in Ref. \cite{Tong-2016-PRA}, $C_{\text{w}}(\rho ,P)$ recovers the standard
coherence measure $C_{\text{w}}(\rho )$ proposed in Ref. \cite{Bu-2018-PRA}, $C_{R,\alpha }(\rho ,P)$\ recovers the standard
coherence measure proposed in Ref. \cite{Xu-2020-CPB}. In particular, when $\alpha =\frac{1}{2},$
\begin{eqnarray}
C_{R,\frac{1}{2}}(\rho ,P)=1-\max_{\sigma \in \mathcal{I}_{\text{B}}(H)}(\text{tr}\sqrt{%
\sqrt{\rho }\sigma \sqrt{\rho }})^{2}   \label{eq15}
\end{eqnarray}
recovers the standard coherence measure proposed in Ref. \cite{Adesso-2015-PRL} when $P$ is a rank-1 projective
measurement.

\section{Coherence measures with respect to general quantum measurements}

We study now the coherence measures with respect to general quantum measurements \cite{Brub-2019-arxiv}.
A general measurement or a POVM on $d$-dimensional Hilbert space $H$ is
given by a set of positive semidefinite operators $E=\{E_{i}%
\}_{i=1}^{n}$ with $\sum_{i=1}^{n}E_{i}=I_{d}$ the identity on $H$. Projective measurement and
rank-1 projective measurement are the special cases of POVM. Suppose $%
E_{i}=A_{i}^{\dagger }A_{i}$ for any $i$. We also denote $%
E=\{A_{i}\}_{i=1}^{n}$ with $\sum_{i=1}^{n}A_{i}^{\dagger }A_{i}=I_{d}$. Note
that $E_{i}=(U_{i}A_{i})^{\dagger }(U_{i}A_{i})$ for any unitary $%
\{U_{i}\}_{i=1}^{n}.$

A state $\rho $ is called an incoherent state with respect to $E$ if
\cite{Brub-2019-PRL}
\begin{eqnarray}
E_{i}\rho E_{j}=0,~~~ \forall i\neq j.  \label{eq16}
\end{eqnarray}
Note that this is equivalent to \cite{Brub-2019-PRL}
\begin{eqnarray}
A_{i}\rho A_{j}^{\dagger }=0, \forall i\neq j.  \label{eq17}
\end{eqnarray}

The POVM incoherent channel is defined via the canonical Naimark extension
\cite{Brub-2019-arxiv}. For POVM $E=\{E_{i}=A_{i}^{%
\dagger }A_{i}\}_{i=1}^{n}$ on $d$-dimensional Hilbert space $H$, introduce
an $n$-dimensional Hilbert space $H_{R}$ with $\{|i\rangle \}_{i=1}^{n}$ an
orthonormal basis of $H_{R}$. A canonical Naimark extension $%
P=\{P_{i}\}_{i=1}^{n}$ of $E=\{E_{i}\}_{i=1}^{n}$ is described by a unitary matrix $V$
on $H_{\varepsilon }=H\otimes H_{R}$ as \cite{Brub-2019-arxiv}
\begin{eqnarray}
&&V=\sum_{ij=1}^{n}A_{ij}\otimes |i\rangle \langle j|,  \label{1-eq1} \\
&&\overline{P}=\{\overline{P}_{i}=I_{d}\otimes |i\rangle \langle
i|\}_{i=1}^{n},  \label{1-eq2}  \\
&&P_{i}=V^{\dagger }\overline{P}_{i}V,   \label{1-eq3}
\end{eqnarray}
with $\{A_{ij}\}_{ij=1}^{n}$ satisfying
\begin{eqnarray}
&&\sum_{i=1}^{n}A_{ij}^{\dagger }A_{ik}=\delta _{jk}I_{d}, \nonumber \\
&&\sum_{k=1}^{n}A_{ik}A_{jk}^{\dagger }=\delta _{ij}I_{d},  \nonumber \\
&&A_{i1}=A_{i}.  \nonumber
\end{eqnarray}
A channel $\phi \in \mathcal{C}(H)$ is called a POVM incoherent (PI) channel
if \cite{Brub-2019-arxiv} $\phi $ allows a Kraus
operator decomposition $\phi =\{K_{l}\}_{l}$ with $\sum_{l}K_{l}^{\dagger
}K_{l}=I_{d}$ and there exists a BI channel $\phi ^{\prime }=\{K_{l}^{\prime
}\}_{l}$ $\in \mathcal{C}_{\text{BI}}(H_{\varepsilon })$ with respect to a
canonical Naimark extension $P=\{P_{i}\}_{i=1}^{n}$ such that
\begin{eqnarray}
K_{l}\rho K_{l}^{\dagger }\otimes |1\rangle \langle 1|=K_{l}^{\prime }(\rho
\otimes |1\rangle \langle 1|)K_{l}^{\prime \dagger }, \ \forall l,
\end{eqnarray}
where $\{K_{l}^{\prime }\}_{l}$ is a BI decomposition of $\phi ^{\prime }.$
For such case we call $\{K_{l}\}_{l}$ a PI decomposition of $\phi .$

We denote the set of all PI states as $\mathcal{I}_{\text{P}}(H),$ and the
set of all PI channels as $\mathcal{C}_{\text{PI}}(H).$ Note that $\mathcal{I%
}_{\text{P}}(H)$ may be empty for some POVMs. Note also that such definition of PI operation does not depend on the choice of Naimark extension \cite{Brub-2019-arxiv},

A coherence measure for states in Hilbert space $H$ with respect to a
general quantum measurement $E=\{E_{i}\}_{i=1}^{n}$ should satisfy the
following conditions (P1)-(P4) \cite{Brub-2019-arxiv}:

\textbf{(P1)} Faithfulness: $C(\rho ,E)\geq 0$,
with equality if $\rho \in \mathcal{I}_{\text{P}}(H)$.

\textbf{(P2)} Monotonicity: $C(\phi _{\text{PI}}(\rho ),E)$ $\leq C(\rho ,E)$, $%
\forall\phi _{\text{PI}}\in \mathcal{C}_{\text{PI}}(H).$

\textbf{(P3)} Strong monotonicity: $\sum_{l}p_{l}C(\rho _{l},P)\leq C(\rho ,P)$, where
$\{K_{l}\}_{l}$ is a PI decomposition of a PI channel, $p_{l}=$tr$(K_{l}\rho
K_{l}^{\dagger })$, $\rho _{l}=K_{l}\rho K_{l}^{\dagger }/p_{l}.$

\textbf{(P4)} Convexity: $C(\sum_{j}p_{j}\rho _{j},E)\leq \sum_{j}p_{j}C(\rho _{j},E)$%
, $\{\rho _{j}\}_{j}\subset \mathcal{S}(H)$, $\{p_{j}\}_{j}$ a probability
distribution.

Note that the definitions of PI states and PI channels and the conditions
(P1)-(P4) all include the projective measurements and the rank-1 projective
measurements as special cases \cite{Brub-2019-arxiv}. We emphasize that the framework of POVM
coherence measure is about POVM $E=\{E_{i}\}_{i=1}^{n}$. Hence, any valid
coherence measure in terms of $\{A_{i}\}_{i}$ should be invariant under the
unitary transformation $\{A_{i}\}_{i}\rightarrow \{U_{i}A_{i}\}_{i}$ for any
unitary $\{U_{i}\}_{i=1}^{n}$ \cite{Brub-2019-arxiv}.

An efficient scheme for constructing POVM coherence measures is as follows
\cite{Brub-2019-PRL,Brub-2019-arxiv}
\begin{eqnarray}
C(\rho ,E)=C(\varepsilon(\rho ),\overline{P}),  \label{eq19}
\end{eqnarray}
where
\begin{eqnarray}
\varepsilon(\rho )=\sum_{ij=1}^{n}A_{i}\rho A_{j}^{\dagger }\otimes
|i\rangle \langle j|,  \label{eq20}
\end{eqnarray}
It can be checked that if $%
C(\rho _{\varepsilon },\overline{P})$ is a unitarily invariant block
coherence measure satisfying conditions (B1) to (B4), then $C(\rho ,E)$
defined above is a POVM coherence measure
satisfying conditions (P1) to (P4) \cite{Brub-2019-arxiv}.
Here $\rho_{\varepsilon }$ is any state on $%
H_{\varepsilon }=H\otimes H_{R}$. The unitary invariance means that
\begin{eqnarray}
C(\rho _{\varepsilon },\overline{P})=C(U\rho _{\varepsilon }U^{\dagger },U%
\overline{P}U^{\dagger })  \label{eq22}
\end{eqnarray}
for any unitary transformation $U$ on $H_{\varepsilon }.$
Employing this scheme and using Propositions 1 to 5, we obtain the following Theorem.

\textbf{Theorem 2.} Let $E=\{E_{i}=A_{i}^{\dagger }A_{i}\}_{i=1}^{n}$ be a POVM on the
Hilbert space $H$. The following quantities given in (1)-(5) are all POVM coherence measures with respect to $E.$

(1). $l_{1}$ norm of coherence
\begin{eqnarray}
 C_{l_{1}}(\rho ,E)=\sum_{i\neq j}||A_{i}\rho A_{j}^{\dagger }||_{\text{tr}}. \label{eq23}
\end{eqnarray}

(2). For $\alpha \in (0,1)\cup (1,2],$ coherence based on Tsallis relative entropy
\begin{eqnarray}
C_{T,\alpha }(\rho ,E)=\frac{1}{\alpha -1}\{\sum_{i}\text{tr}[(A_{i}\rho
^{\alpha }A_{i}^{\dagger })^{1/\alpha }]-1\},  \label{eq24}
\end{eqnarray}
and
\begin{eqnarray}
\lim_{\alpha \rightarrow 1}C_{T,\alpha }(\rho ,E)=(\ln 2)C_{\text{rel}%
}(\rho ,E),  \label{eq25}
\end{eqnarray}
where
\begin{eqnarray}
C_{\text{rel}}(\rho ,E)=\text{tr}(\rho \log _{2}\rho )-\sum_{i}\text{tr}[(A_{i}\rho
A_{i}^{\dagger })\log _{2}(A_{i}\rho A_{i}^{\dagger })].  \label{eq26} \nonumber \\
\end{eqnarray}

(3).  Modified trace norm of coherence
\begin{eqnarray}
C_{\text{tr}}(\rho ,E)=\min_{\lambda >0,\sigma \in \mathcal{I}_{\text{B%
}}(H_{\varepsilon })}||\varepsilon(\rho )-\lambda \sigma ||_{\text{tr}%
}.  \label{eq27}
\end{eqnarray}

(4).  Coherence weight
\begin{eqnarray}
C_{\text{w}}(\rho ,E)=\min_{\sigma \in \mathcal{I}_{\text{B}}(H_{\varepsilon
})}\{s\geq 0|\varepsilon(\rho )\geq (1-s)\sigma \}.  \label{eq28}
\end{eqnarray}

(5). For $\alpha \in \lbrack \frac{1}{2},1),$ coherence based on sandwiched R\'{e}nyi relative entropy
\begin{eqnarray}
&&C_{R,\alpha }(\rho ,E)  \nonumber \\
&=&1-\max_{\sigma \in \mathcal{I}_{\text{B}%
}(H_{\varepsilon })}\{\text{tr}[(\varepsilon(\rho ^{\frac{1-\alpha }{%
2\alpha }})\sigma \varepsilon(\rho ^{\frac{1-\alpha }{2\alpha }%
}))^{\alpha }]\}^{\frac{1}{1-\alpha }}. \ \  \label{eq29}
\end{eqnarray}

{\sf [Proof].}
To prove the results of the Theorem 2, we need to use the results of the Propositions 1 to 5.
Let $\{|i\rangle \}_{i=1}^{n}$ be an orthonormal basis
for the Hilbert space $H_{R}$, and $\overline{P}$ and $\varepsilon(\rho )$ be defined in Eqs. (\ref{1-eq2}) and (\ref{eq20}), respectively.
Since $C(\rho ,E)$ is a POVM coherence measure satisfying conditions (P1) to (P4)
if $C(\rho _{\varepsilon },\overline{P})$ is a unitarily invariant block
coherence measure satisfying conditions (B1) to (B4),
we only need to prove the unitary invariance Eq. (\ref{eq22})
and show that $C_{l_{1}}(\rho ,E)$, $C_{T,\alpha }(\rho ,E)$, $C_{\text{tr}}(\rho ,E)$,
$C_{\text{w}}(\rho ,E)$ and $C_{R,\alpha }(\rho ,E)$ take the forms of Eqs. (\ref{eq23}), (\ref{eq24}), (\ref{eq27}), (\ref{eq28}) and (\ref{eq29}) under Eq. (\ref{eq19}), respectively.

(1). We prove that $C_{l_{1}}(\rho _{\varepsilon },\overline{P})$ is
unitarily invariant. For any unitary $U$ on $H_{\varepsilon }$, we have
\begin{eqnarray}
&&C_{l_{1}}(U\rho _{\varepsilon }U^{\dagger },U\overline{P}U^{\dagger }) \nonumber \\
&=&\sum_{i\neq j}||U\overline{P}_{i}U^{\dagger }U\rho _{\varepsilon
}U^{\dagger }U\overline{P}_{j}U^{\dagger }||_{\text{tr}} \nonumber \\
&=&\sum_{i\neq j}||\overline{P}_{i}\rho _{\varepsilon }\overline{P}_{j}||_{%
\text{tr}}
=C_{l_{1}}(\rho _{\varepsilon },\overline{P}), \label{eqA77}  \nonumber
\end{eqnarray}
where we have used the fact that the trace norm is unitarily invariant. It
is easy to see that $C_{l_{1}}(\rho ,E)$ have the form of Eq. (\ref{eq23}).

(2). It is easy to see that $C_{T,\alpha }(\rho _{\varepsilon },\overline{P}%
) $ is unitarily invariant. Now we prove that $C_{T,\alpha }(\rho ,E)$ has
the form of Eq. (\ref{eq24}) under Eq. (\ref{eq19}).

For the unitary transformation $V$ defined in Eq. (\ref{1-eq1}),
\begin{eqnarray}
\varepsilon _{V}(\rho )=V(\rho \otimes |1\rangle \langle 1|)V^{\dagger
}=\sum_{ij}A_{i}\rho A_{j}^{\dagger }\otimes |i\rangle \langle j|=\varepsilon(\rho ). \label{eqA81} \nonumber
\end{eqnarray}
As a result,
\begin{eqnarray}
&&\text{tr}[(\overline{P}_{i}(\varepsilon _{V}(\rho ))^{\alpha }\overline{P}%
_{i})^{1/\alpha }] \nonumber \\
&=&\text{tr}[(\overline{P}_{i}V(\rho ^{\alpha }\otimes |1\rangle \langle
1|)V_{i}^{\dagger }\overline{P})^{1/\alpha }] \nonumber \\
&=&\text{tr}[(\overline{P}_{i}(\sum_{jk}A_{j}\rho ^{\alpha }A_{k}^{\dagger }\otimes
|j\rangle \langle k|)\overline{P}_{i})^{1/\alpha }] \nonumber \\
&=&\text{tr}[(A_{i}\rho A_{i}^{\dagger }\otimes |i\rangle \langle i|)^{1/\alpha }] \nonumber \\
&=&\text{tr}[(A_{i}\rho A_{i}^{\dagger })^{1/\alpha }]. \label{eqA82} \nonumber
\end{eqnarray}
Hence, $C_{T,\alpha }(\rho ,E)$ has the form of Eq. (\ref{eq24}).
Eq. (\ref{eq25}) can be proved as Corollary 1.

(3). It is easy to see that $C_{\text{tr}}(\rho,E)$ has the form of Eq.
(\ref{eq27}). Now we show that $C_{\text{tr}}(\rho _{\varepsilon },\overline{P})$ is
unitarily invariant. Note that
\begin{eqnarray}
C_{\text{tr}}(\rho _{\varepsilon },\overline{P})=\min_{\lambda >0,\sigma
}||\rho _{\varepsilon }-\lambda \sum_{i=1}^{n}\overline{P}_{i}\sigma
\overline{P}_{i}||_{\text{tr}}, \label{eqA83} \nonumber
\end{eqnarray}
where $\sigma $ is any density operator on $H_{\varepsilon }.$

For any unitary $U$ on $H_{\varepsilon }$, we have
\begin{eqnarray}
&&C_{\text{tr}}(U\rho _{\varepsilon }U^{\dagger },U\overline{P}U^{\dagger
})   \nonumber \\
&=&\min_{\lambda >0,\sigma }||U\rho _{\varepsilon }U^{\dagger }-\lambda
\sum_{i=1}^{n}U\overline{P}_{i}U^{\dagger }\sigma U\overline{P}%
_{i}U^{\dagger }||_{\text{tr}} \nonumber \\
&=&\min_{\lambda >0,\sigma }||\rho _{\varepsilon }-\lambda \sum_{i=1}^{n}%
\overline{P}_{i}U^{\dagger }\sigma U\overline{P}_{i}||_{\text{tr}} \nonumber \\
&=&\min_{\lambda >0,\sigma }||\rho _{\varepsilon }-\lambda \sum_{i=1}^{n}%
\overline{P}_{i}\sigma \overline{P}_{i}||_{\text{tr}} \nonumber \\
&=&C_{\text{tr}}(\rho _{\varepsilon },\overline{P}),  \label{eqA84} \nonumber
\end{eqnarray}
where we have used the facts that trace norm is unitarily invariant and $%
\{\sigma :\sigma \in \mathcal{S}(H)\}=\{U^{\dagger }\sigma U:\sigma \in
\mathcal{S}(H)\}$.

(4). It is easy to see that $C_{\text{w}}(\rho ,E)$ has the form of Eq. (\ref{eq28}).
Next we show that $C_{\text{w}}(\rho _{\varepsilon },\overline{P})$ is
unitarily invariant. Note that
\begin{eqnarray}
C_{\text{w}}(\rho _{\varepsilon },\overline{P})=\min_{\sigma }\{s\geq
0|\rho _{\varepsilon }\geq (1-s)\sum_{i=1}^{n}\overline{P}_{i}\sigma
\overline{P}_{i}\}, \ \  \label{eqA85} \nonumber
\end{eqnarray}
where $\sigma $ is any density operator on $H_{\varepsilon }.$

For any unitary $U$ on $H_{\varepsilon }$, we have
\begin{eqnarray}
&&C_{\text{w}}(U\rho _{\varepsilon }U^{\dagger },U\overline{P}U^{\dagger })  \nonumber \\
&=&\min_{\sigma }\{s\geq 0|U\rho _{\varepsilon }U^{\dagger }\geq
(1-s)\sum_{i=1}^{n}U\overline{P}_{i}U^{\dagger }\sigma U\overline{P}%
_{i}U^{\dagger }\}  \nonumber \\
&=&\min_{\sigma }\{s\geq 0|\rho _{\varepsilon }\geq (1-s)\sum_{i=1}^{n}%
\overline{P}U^{\dagger }\sigma U\overline{P}_{i}\}  \nonumber \\
&=&\min_{\sigma }\{s\geq 0|\rho _{\varepsilon }\geq (1-s)\sum_{i=1}^{n}%
\overline{P}_{i}\sigma \overline{P}_{i}\} \nonumber \\
&=&C_{\text{w}}(\rho _{\varepsilon },\overline{P}),   \label{eqA86} \nonumber
\end{eqnarray}
which completes the proof.

(5). It is easy to see that $C_{R,\alpha }(\rho ,E)$ has the form of Eq.
(\ref{eq29}). Similarly to the proof of (3), one can show that $C_{\text{w}}(\rho _{\varepsilon },\overline{P})$ is unitarily invariant.
\hfill \rule{1ex}{1ex}

We remark that the coherence measure $C_{l_{1}}(\rho ,P)$ was proposed in \cite{Aberg-2006-arxiv}.
In \cite{Brub-2019-arxiv} the authors conjectured that $C_{l_{1}}(\rho ,E)$ is a well defined POVM coherence measure satisfying the conditions (P1)-(P4). Combining with our result of proposition 1, we have strictly proved in Theorem 2 that $C_{l_{1}}(\rho ,E)$ is indeed a well defined POVM coherence measure.

\section{Summary}

We have established an alternative framework for quantifying the
coherence with respect to projective measurements, and provided several
coherence measures with respect to projective measurements.
We then obtained several coherence measures with respect to general POVM measurements, from which
a conjecture has been verified concerning the coherence measure $C_{l_{1}}(\rho ,E)$.
The coherence with respect to POVM measurements has operational significance.
Our results may highlight further investigations on the coherence of quantum states
and the applications in quantum information processing.

\section*{ACKNOWLEDGMENTS}
This work was supported by the National Science Foundation of China (Grant No. 11675113), the Key Project of Beijing Municipal Commission of Education (Grant No. KZ201810028042), Beijing Natural Science Foundation (Grant No. Z190005), and Young Talent fund of University Association for Science and Technology
in Shaanxi (Grant No. 20190111).

\section*{Appendix}
\setcounter{equation}{0} \renewcommand%
\theequation{A\arabic{equation}}

\subsection{Proof of Proposition 1}

From the definition of BI state and the properties of trace norm, $C_{l_{1}}(\rho ,P)$ satisfies
the condition (B1). It satisfies the conditions (B4)
and (B5) due to the properties of trace norm. Since (B3) and (B4) imply (B2), we only need to prove
that $C_{l_{1}}(\rho ,P)$ fulfills (B3).

For any BI channel $\phi$ with BI decomposition $\phi_{\text{BI}}=\{K_{l}\}_{l}$,
$\sum_{l}K_{l}^{\dagger }K_{l}=I_{d}$,
each $K_{l}$ has the form \cite{Brub-2019-arxiv},
\begin{eqnarray}
K_{l}=\sum_{i=1}^{n}P_{f_{l}(i)}M_{l}P_{i},  \label{eqA11}
\end{eqnarray}
where $f_{l}(i)$ is a function on $\{i\}_{i=1}^{n},$ $M_{l}$ is a matrix on $%
H$.
Denote $p_{l}=$tr$(K_{l}\rho K_{l}^{\dagger })$, $\rho _{l}=K_{l}\rho
K_{l}^{\dagger }/p_{l}$. We have
\begin{widetext}
\begin{eqnarray}
&&\sum_{l}p_{l}C_{l_{1}}(\rho _{l},P)
=\sum_{l,i\neq j}||P_{i}K_{l}\rho K_{l}^{\dagger }P_{j}||_{\text{tr}}   \label{eqA13}  \nonumber \\
&=&\sum_{l,i\neq j}||P_{i}K_{l}\sum_{i^{\prime }\neq j^{\prime }}P_{i^{\prime
}}\rho P_{j^{\prime }}K_{l}^{\dagger }P_{j}||_{\text{tr}}    \label{eqA14} \\
&\leq&\sum_{l,ij,i^{\prime }\neq j^{\prime }}||P_{i}K_{l}P_{i^{\prime }}\rho
P_{j^{\prime }}K_{l}^{\dagger }P_{j}||_{\text{tr}}   \label{eqA15}  \nonumber \\
&=&\sum_{l,i^{\prime }\neq j^{\prime }}||P_{f_{l}(i^{\prime
})}K_{l}P_{i^{\prime }}\rho P_{j^{\prime }}K_{l}^{\dagger
}P_{f_{l}(j^{\prime })}||_{\text{tr}}     \label{eqA16} \\
&=&\sum_{l,i^{\prime }\neq j^{\prime }}||P_{f_{l}(i^{\prime
})}K_{l}\sum_{k}s_{i^{\prime }j^{\prime }k}|\psi _{i^{\prime }j^{\prime
}k}\rangle \langle \overline{\psi }_{i^{\prime }j^{\prime }k}|K_{l}^{\dagger
}P_{f_{l}(j^{\prime })}||_{\text{tr}}   \label{eqA17} \\
&\leq&\sum_{lk,i^{\prime }\neq j^{\prime }}s_{i^{\prime }j^{\prime
}k}||P_{f_{l}(i^{\prime })}K_{l}|\psi _{i^{\prime }j^{\prime }k}\rangle
\langle \overline{\psi }_{i^{\prime }j^{\prime }k}|K_{l}^{\dagger
}P_{f_{l}(j^{\prime })}||_{\text{tr}}    \label{eqA18}  \nonumber \\
&=&\sum_{k,i^{\prime }\neq j^{\prime }}s_{i^{\prime }j^{\prime }k}\sum_{l}\sqrt{%
\langle \psi _{i^{\prime }j^{\prime }k}|K_{l}^{\dagger }P_{f_{l}(i^{\prime
})}K_{l}|\psi _{i^{\prime }j^{\prime }k}\rangle \langle \overline{\psi }%
_{i^{\prime }j^{\prime }k}|K_{l}^{\dagger }P_{f_{l}(j^{\prime })}K_{l}|%
\overline{\psi }_{i^{\prime }j^{\prime }k}\rangle }  \label{eqA19} \\
&\leq&\sum_{k,i^{\prime }\neq j^{\prime }}s_{i^{\prime }j^{\prime
}k}\sqrt{\sum_{l}\langle \psi _{i^{\prime }j^{\prime }k}|K_{l}^{\dagger
}P_{f_{l}(i^{\prime })}K_{l}|\psi _{i^{\prime }j^{\prime }k}\rangle}\sqrt{
\sum_{l^{\prime }}\langle \overline{\psi }_{i^{\prime }j^{\prime }k}|K_{l^{\prime }}^{\dagger }P_{f_{l^{\prime }}(j^{\prime })}K_{l^{\prime }}|%
\overline{\psi }_{i^{\prime }j^{\prime }k}\rangle} \label{eqA20}   \\
&=&\sum_{k,i^{\prime }\neq j^{\prime }}s_{i^{\prime }j^{\prime }k}\sqrt{\langle
\psi _{i^{\prime }j^{\prime }k}|\sum_{l}K_{l}^{\dagger }P_{f_{l}(i^{\prime
})}K_{l}|\psi _{i^{\prime }j^{\prime }k}\rangle}\sqrt{\langle \overline{\psi }%
_{i^{\prime }j^{\prime }k}|\sum_{l^{\prime }}K_{l^{\prime
}}^{\dagger }P_{f_{l^{\prime }}(j^{\prime })}K_{l^{\prime }}|\overline{\psi }%
_{i^{\prime }j^{\prime }k}\rangle}    \label{eqA21}  \nonumber \\
&\leq&\sum_{k,i^{\prime }\neq j^{\prime }}s_{i^{\prime }j^{\prime }k}\sqrt{\langle
\psi _{i^{\prime }j^{\prime }k}|I_{m}|\psi _{i^{\prime }j^{\prime }k}\rangle}
\sqrt{\langle \overline{\psi }_{i^{\prime }j^{\prime }k}|I_{m}|\overline{%
\psi }_{i^{\prime }j^{\prime }k}\rangle}  \label{eqA22} \\
&=&\sum_{k,i^{\prime }\neq j^{\prime }}s_{i^{\prime }j^{\prime }k}
=\sum_{i^{\prime }\neq j^{\prime }}||P_{i^{\prime }}\rho P_{j^{\prime
}}||_{\text{tr}}=C_{l_{1}}(\rho ,P). \label{eqA23} \nonumber
\end{eqnarray}
\end{widetext}
In Eq. \eqref{eqA14} we have used the property that $\{K_{l}\}_{l}$ is a BI decomposition,
that is, $P_{i}K_{l}(\sum_{i^{\prime }}P_{i^{\prime }}\rho P_{i^{\prime
}})K_{l}^{\dagger }P_{j}=0$ for any $i\neq j.$ In Eq.  \eqref{eqA16}  we have used $%
P_{i}K_{l}P_{i^{\prime }}=P_{i}P_{f_{l}(i^{\prime })}K_{l}P_{i^{\prime }}=\delta_{i,f_{l}(i^{\prime })}P_{f_{l}(i^{\prime })}K_{l}P_{i^{\prime }}$ which
is a result of Eq.  \eqref{eqA11}. In Eq.  \eqref{eqA17}  we have used the singular value decomposition, $%
P_{i^{\prime }}\rho P_{j^{\prime }}=\sum_{k}s_{i^{\prime }j^{\prime }k}|\psi
_{i^{\prime }j^{\prime }k}\rangle \langle \overline{\psi }_{i^{\prime
}j^{\prime }k}|$ with $\{s_{i^{\prime }j^{\prime }k}\}_{k}$ the singular
values, $\{|\psi _{i^{\prime }j^{\prime }k}\rangle \}_{k}$ ($\{|\overline{\psi }_{i^{\prime }j^{\prime
}k}\rangle \}_{k}$) a set of orthonormal vectors. In Eq. \eqref{eqA19} we have taken into account
the fact that $|||\psi \rangle \langle \varphi |||_{\text{tr}}=\sqrt{\langle \psi |\psi
\rangle \langle \varphi |\varphi \rangle }$ for any pure states $|\psi
\rangle $ and $|\varphi \rangle .$ In Eq.  \eqref{eqA20} we have used the Cauchy-Schwarz
inequality $\sum_{l}\sqrt{a_{l}b_{l}}\leq\sqrt{\sum_{l}a_{l}}\sqrt{\sum_{l^{\prime }}b_{l^{\prime }}}$ with $a_{l}\geq0$ and $b_{l}\geq0$. In Eq.  \eqref{eqA22} we have used the fact that $\sum_{l}K_{l}^{\dagger
}P_{f_{l}(i^{\prime })}K_{l}\leq I_{m}$ since $P_{f_{l}(i^{\prime })}\leq I_{m}$ and
$\sum_{l}K_{l}^{\dagger }K_{l}=I_{m}.$

\subsection{Proof of Proposition 2}

For $\alpha >0,$ the quantum Tsallis relative entropy is
defined as
\begin{eqnarray}
D_{T,\alpha }(\rho ||\sigma ) &=&\frac{\text{tr}(\rho ^{\alpha }\sigma ^{1-\alpha
})-1}{\alpha -1},~~\rho,\,\sigma \in \mathcal{S(}H\mathcal{)},  \nonumber \\
\ \text{supp}(\rho ) &\subset &\text{supp}(\sigma )\ \text{when}\ \alpha
\geq 1,  \label{eqA24}
\end{eqnarray}%
where supp$(\rho )=\{|\psi \rangle |\rho |\psi \rangle \neq 0\}$ is the
support of $\rho .$

It is shown that for $\alpha >0$ \cite{Rastegin-2016-PRA},
\begin{eqnarray}
D_{T,\alpha }(\rho ||\sigma )\geq 0, \ \ D_{T,\alpha }(\rho ||\sigma
)=0\Leftrightarrow \rho =\sigma.  \label{eqA25}
\end{eqnarray}
Also, $D_{\alpha }(\rho ||\sigma )$ is monotonic under CPTP maps for $\alpha
\in (0;2]$ \cite{Rastegin-2016-PRA},
\begin{eqnarray}
D_{T,\alpha }(\phi (\rho )||\phi (\sigma ))\leq D_{T,\alpha }(\rho ||\sigma
).  \label{eqA26}
\end{eqnarray}
Define
\begin{eqnarray}
D_{T,\alpha }(\rho )=\min_{\sigma \in \mathcal{I}_{\text{B}}(H)}D_{T,\alpha }(\rho
||\sigma ).  \label{eqA27}
\end{eqnarray}
We now prove that
\begin{eqnarray}
D_{T,\alpha }(\rho )=\frac{\{\sum_{i}\text{tr}[(P_{i}\rho ^{\alpha }P_{i})^{\frac{1%
}{\alpha }}]\}^{\alpha }-1}{\alpha -1}.  \label{eqA28}
\end{eqnarray}
To go ahead, we need the lemmas below.

\textbf{Lemma 1.} H\"{o}lder inequality.

Suppose $\{a_{i}\}_{i=1}^{n},\{b_{i}\}_{i=1}^{n},$ are all positive real
numbers, then

1) when $\alpha \in (0,1),$
\begin{eqnarray}
\sum_{i=1}^{n}a_{i}b_{i}\leq (\sum_{i=1}^{n}a_{i}^{\frac{1}{\alpha }%
})^{\alpha }(\sum_{i=1}^{n}b_{i}^{\frac{1}{1-\alpha }})^{1-\alpha }, \label{eqA29}
\end{eqnarray}
and the equality holds if and only if $a_{i}^{\frac{1}{\alpha }}/b_{i}^{\frac{1}{%
1-\alpha }}=a_{j}^{\frac{1}{\alpha }}/b_{j}^{\frac{1}{1-\alpha }}$ for any $%
i,j;$

2) when $\alpha >1,$
\begin{eqnarray}
\sum_{i=1}^{n}a_{i}b_{i}\geq (\sum_{i=1}^{n}a_{i}^{\frac{1}{\alpha }%
})^{\alpha }(\sum_{i=1}^{n}b_{i}^{\frac{1}{1-\alpha }})^{1-\alpha }, \label{eqA30}
\end{eqnarray}
and the equality holds if and only if $a_{i}^{\frac{1}{\alpha }}/b_{i}^{\frac{1}{%
1-\alpha }}=a_{j}^{\frac{1}{\alpha }}/b_{j}^{\frac{1}{1-\alpha }}$ for any $%
i,j.$

\textbf{Lemma 2} (Ref. \cite{Marshall-2011-book}).
For $r\times r$ positive semidefinite matrices $M$ and $N$, it holds that
\begin{eqnarray}
\sum_{j=1}^{r}\lambda _{r+1-j}^{\downarrow }(M)\lambda _{j}^{\downarrow
}(N)\leq \text{tr}(MN)\leq \sum_{j=1}^{r}\lambda _{j}^{\downarrow }(M)\lambda
_{j}^{\downarrow }(N), \nonumber \\  \label{eqA31}
\end{eqnarray}
where $\{\lambda _{j}^{\downarrow }(M)\}_{j}$ are the eigenvalues of $M$ in
decreasing order.

Now for $\alpha \in (0,1)$ and $\sigma \in \mathcal{I}_{\text{B}}(H)$, we have
\begin{eqnarray}
&&\text{tr}(\rho ^{\alpha }\sigma ^{1-\alpha })  \nonumber \\
&=&\text{tr}[\rho ^{\alpha }\sum_{i=1}^{n}(P_{i}\sigma P_{i})^{1-\alpha }]   \nonumber \\
&=&\sum_{i=1}^{n}q_{i}^{1-\alpha }\text{tr}(\rho ^{\alpha }\sigma _{i}^{1-\alpha
})\leq \{\sum_{i=1}^{n}[\text{tr}(\rho ^{\alpha }\sigma _{i}^{1-\alpha })]^{\frac{1}{\alpha }}\}^{\alpha },  \nonumber \\  \label{eqA32}
\end{eqnarray}
where $q_{i}=$tr$(P_{i}\sigma P_{i})$, $\sigma_{i}=P_{i}\sigma P_{i}/q_{i}$,
the H\"{o}lder inequality has been used, and
the equality holds if and only if there exists constant $C\geq 0$ such that $q_{i}=C[$tr$%
(\rho ^{\alpha }\sigma _{i}^{1-\alpha })]^{\frac{1}{\alpha }}$ for any $i.$
Furthermore,
\begin{eqnarray}
&&\text{tr}(\rho ^{\alpha }\sigma _{i}^{1-\alpha })  \nonumber \\
&=&\text{tr}(\rho ^{\alpha }P_{i}\sigma _{i}^{1-\alpha }P_{i})  \nonumber \\
&\leq&\sum_{j=1}^{m_{i}}\lambda _{j}^{\downarrow }(P_{i}\rho ^{\alpha
}P_{i})\lambda _{j}^{\downarrow }(\sigma _{i}^{1-\alpha })  \nonumber \\
&=&\sum_{j=1}^{m_{i}}\lambda _{j}^{\downarrow }(P_{i}\rho ^{\alpha
}P_{i})(\lambda _{j}^{\downarrow }(\sigma _{i}))^{1-\alpha }  \nonumber \\
&\leq&\{\sum_{j=1}^{m_{i}}[\lambda _{j}^{\downarrow }(P_{i}\rho ^{\alpha
}P_{i})]^{\frac{1}{\alpha }}\}^{\alpha }\{\sum_{j=1}^{m_{i}}[(\lambda
_{j}^{\downarrow }(\sigma _{i}))^{1-\alpha }]^{\frac{1}{1-\alpha }%
}\}^{1-\alpha }  \nonumber \\
&=&\{\text{tr}[(P_{i}\rho ^{\alpha }P_{i})^{\frac{1}{\alpha }}]\}^{\alpha }, \label{eqA33}
\end{eqnarray}
where the Lemma 1 and Lemma 2 have been used. It is easy to check
that when
\begin{eqnarray}
\sigma =\frac{\sum_{i=1}^{n}(P_{i}\rho ^{\alpha }P_{i})^{\frac{1}{\alpha }}}{\sum_{i=1}^{n}\text{tr}[(P_{i}\rho ^{\alpha }P_{i})^{\frac{1}{\alpha }}]}  \label{1-eqA1}
\end{eqnarray}
Eq. \eqref{eqA27} achieves Eq. \eqref{eqA28}. As a result we get Eq. \eqref{eqA28}.

For $\alpha >1,$ we have
\begin{eqnarray}
&&\text{tr}(\rho ^{\alpha }\sigma ^{1-\alpha })   \nonumber \\
&=&\text{tr}[\rho ^{\alpha }\sum_{i}(P_{i}\sigma P_{i})^{1-\alpha }]   \nonumber \\
&=&\sum_{i}q_{i}^{1-\alpha }\text{tr}(\rho ^{\alpha }\sigma _{i}^{1-\alpha })  \nonumber \\
&\geq&\{\sum_{i}[\text{tr}(\rho ^{\alpha }\sigma _{i}^{1-\alpha })]^{\frac{1}{%
\alpha }}\}^{\alpha }, \label{eqA34}
\end{eqnarray}
and the equality holds if and only if there exists a constant $C\geq 0$ such that $q_{i}=C[$tr$%
(\rho ^{\alpha }\sigma _{i}^{1-\alpha })]^{\frac{1}{\alpha }}$ for any $i.$
Moreover,
\begin{eqnarray}
&&\text{tr}(\rho ^{\alpha }\sigma _{i}^{1-\alpha })  \nonumber \\
&=&\text{tr}(\rho ^{\alpha }P_{i}\sigma _{i}^{1-\alpha }P_{i})  \nonumber \\
&\geq&\sum_{j=1}^{m_{i}}\lambda _{j}^{\downarrow }(P_{i}\rho ^{\alpha
}P_{i})\lambda _{m_{i}+1-j}^{\downarrow }(\sigma _{i}^{1-\alpha })  \nonumber \\
&=&\sum_{j=1}^{m_{i}}\lambda _{j}^{\downarrow }(P_{i}\rho ^{\alpha
}P_{i})(\lambda _{m_{i}+1-j}^{\downarrow }(\sigma _{i}))^{1-\alpha }   \nonumber \\
&\geq&\{\sum_{j=1}^{m_{i}}[\lambda _{j}^{\downarrow }(P_{i}\rho ^{\alpha
}P_{i})]^{\frac{1}{\alpha }}\}^{\alpha }\{\sum_{j=1}^{m_{i}}[(\lambda
_{m_{i}+1-j}^{\downarrow }(\sigma _{i}))^{1-\alpha }]^{\frac{1}{1-\alpha }%
}\}^{1-\alpha }   \nonumber \\
&=&\{\text{tr}[(P_{i}\rho ^{\alpha }P_{i})^{\frac{1}{\alpha }}]\}^{\alpha }. \label{eqA35}
\end{eqnarray}
In above derivation, we have used Lemma 1 and Lemma 2. Again, when $\sigma$ takes the value in Eq. \eqref{1-eqA1}, Eq. \eqref{eqA27} achieves Eq. \eqref{eqA28}. As a result we get Eq. \eqref{eqA28}.

From Eqs. (\ref{eqA25}) and (\ref{eqA27}) we see that $D_{T,\alpha }(\rho )\geq 0$ and $D_{T,\alpha }(\rho
)=0$ if and only if $\rho \in \mathcal{I}_{\text{B}}(H)$. Then from Eq. (\ref{eqA28}) we have
\begin{eqnarray}
\frac{\{\sum_{i}\text{tr}[(P_{i}\rho ^{\alpha }P_{i})^{\frac{1}{\alpha }%
}]\}^{\alpha }-1}{\alpha -1}\geq 0, \nonumber \label{eqA36}
\end{eqnarray}
namely,
\begin{eqnarray}
\frac{\sum_{i}\text{tr}[(P_{i}\rho ^{\alpha }P_{i})^{\frac{1}{\alpha }}]-1}{\alpha
-1}\geq 0,\nonumber \label{eqA37}
\end{eqnarray}
with the equality holding if and only if $\rho \in \mathcal{I}_{\text{B}}(H)$,
which proves that $C_{T,\alpha }(\rho ,P)$ satisfies (B1).

For any $\phi _{\text{BI}}\in \mathcal{C}_{\text{BI}}(H)$, from Eqs. (\ref{eqA26}) and (\ref{eqA27}) we have
\begin{eqnarray}
&&D_{T,\alpha }(\rho )=\min_{\sigma \in \mathcal{I}_{\text{B}}(H)}D_{T,\alpha }(\rho
||\sigma )
=D_{T,\alpha }(\rho ||\sigma ^{\ast })   \nonumber \\
&\geq&D_{T,\alpha }(\phi _{\text{BI}}(\rho )||\phi _{\text{BI}}(\sigma
^{\ast }))   \nonumber \\
&\geq&\min_{\sigma \in \mathcal{I}_{\text{B}}(H)}D_{T,\alpha }(\phi _{\text{%
BI}}(\rho )||\sigma )
=D_{T,\alpha }(\phi _{\text{BI}}(\rho )),  \label{eqA38}
\end{eqnarray}
where $\sigma ^{\ast }\in \mathcal{I}_{\text{B}}(H)$ such that $\min_{\sigma \in
\mathcal{I}_{\text{B}}(H)}D_{T,\alpha }(\rho ||\sigma )=D_{T,\alpha }(\rho ||\sigma
^{\ast }).$

From Eq.  \eqref{eqA28}, Eq.  \eqref{eqA38} is equivalent to
\begin{eqnarray}
&&\frac{\{\sum_{i}\text{tr}[(P_{i}\rho ^{\alpha }P_{i})^{\frac{1}{\alpha }%
}]\}^{\alpha }-1}{\alpha -1} \nonumber \\
&\leq&\frac{\{\sum_{i}\text{tr}[(P_{i}^{\alpha }(\phi
_{\text{BI}}(\rho ))^{\alpha }P_{i})^{\frac{1}{\alpha }}]\}^{\alpha }-1}{\alpha -1}%
,  \label{eqA39} \nonumber
\end{eqnarray}
which is further equivalent to
\begin{eqnarray}
&&\frac{\sum_{i}\text{tr}[(P_{i}\rho ^{\alpha }P_{i})^{\frac{1}{\alpha }}]-1}{\alpha
-1}  \nonumber \\
&\leq&\frac{\sum_{i}\text{tr}[(P_{i}^{\alpha }(\phi _{\text{BI}}(\rho ))^{\alpha
}P_{i})^{\frac{1}{\alpha }}]-1}{\alpha -1}.  \label{eqA40} \nonumber
\end{eqnarray}
We then proved that $C_{T,\alpha }(\rho ,P)$ satisfies (B2).

Now we prove that $C_{T,\alpha }(\rho ,P)$ also satisfies (B5). Suppose
$\rho =p_{1}\rho _{1}\oplus p_{2}\rho _{2}$ as described in (B5). Then
\begin{eqnarray}
&&\sum_{i=1}^{n}\text{tr}[(P_{i}\rho ^{\alpha }P_{i})^{\frac{1}{\alpha }}]   \nonumber \\
&=&p_{1}\sum_{k_{1}}\text{tr}[(P_{k_{1}}\rho _{1}^{\alpha }P_{k_{1}})^{\frac{1}{%
\alpha }}]+p_{2}\sum_{k_{2}}\text{tr}[(P_{k_{2}}\rho _{2}^{\alpha }P_{k_{2}})^{%
\frac{1}{\alpha }}]   \nonumber \\
&=&p_{1}\sum_{i=1}^{n}\text{tr}[(P_{i}\rho _{1}^{\alpha }P_{i})^{\frac{1}{\alpha }%
}]+p_{2}\sum_{i=1}^{n}\text{tr}[(P_{i}\rho _{2}^{\alpha }P_{i})^{\frac{1}{\alpha }%
}].  \label{eqA42}
\end{eqnarray}
Substituting \eqref{eqA42} into Eq. \eqref{eq9}, we then proved that $C_{T,\alpha }(\rho ,P)$
satisfies (B5).

\subsection{Proof of Corollary 1}

Set $\alpha =1+\varepsilon $. Consider the Taylor expansions around $%
\varepsilon =0,$
\begin{eqnarray}
M^{1+\varepsilon }&=&M+\varepsilon M\ln M+o(\varepsilon ^{2}),   \label{eqA43}   \nonumber \\
\ln (M+\varepsilon N)&=&\ln M+o(\varepsilon ),    \label{eqA44}   \nonumber \\
\frac{1}{1+\varepsilon }&=&1-\varepsilon +o(\varepsilon ^{2}),   \label{eqA45} \nonumber
\end{eqnarray}
where $M$, $N$ are Hermitian matrices, $o(\varepsilon )$ denotes the
infinitesimal term with the order $\varepsilon $ or higher around $%
\varepsilon =0.$
We have $P_{i}\rho ^{\alpha }P_{i}=P_{i}(\rho +\varepsilon \rho \ln \rho
+o(\varepsilon ^{2}))P_{i}$. Therefore,
\begin{eqnarray}
&&\text{tr}[(P_{i}\rho ^{\alpha }P_{i})^{\frac{1}{\alpha }}] \nonumber \\
&=&\texttt{tr}[(P_{i}\rho ^{\alpha }P_{i})^{1-\varepsilon +o(\varepsilon ^{2})}] \nonumber \\
&=&\text{tr}[(P_{i}\rho ^{\alpha }P_{i})-\varepsilon (P_{i}\rho ^{\alpha
}P_{i})\ln (P_{i}\rho ^{\alpha }P_{i})+o(\varepsilon ^{2})]  \nonumber \\
&=&\text{tr}[P_{i}\rho P_{i}+\varepsilon P_{i}(\rho \ln \rho )P_{i}-\varepsilon
(P_{i}\rho P_{i})\ln (P_{i}\rho P_{i})  \nonumber \\
&& \ \ \ \   +o(\varepsilon ^{2})].  \label{eqA47} \nonumber
\end{eqnarray}
Applying the L'Hospital's rule to Eq. \eqref{eq9}, we have
\begin{eqnarray}
&&\lim_{\alpha \rightarrow 1}C_{T,\alpha }(\rho ,P)  \nonumber \\
&=&\lim_{\alpha \rightarrow 1}\frac{d}{d\alpha }\sum_{i}\text{tr}[(P_{i}\rho
^{\alpha }P_{i})^{1/\alpha }]  \nonumber \\
&=&\sum_{i}\text{tr}[P_{i}(\rho \ln \rho )P_{i}-(P_{i}\rho P_{i})\ln (P_{i}\rho
P_{i})]  \nonumber \\
&=&\text{tr}(\rho \ln \rho )-\sum_{i}\text{tr}[(P_{i}\rho P_{i})\ln (P_{i}\rho P_{i})] \nonumber \\
&=&(\ln 2)C_{\text{rel}}(\rho,P).  \label{eqA48} \nonumber
\end{eqnarray}

\subsection{Proof of Proposition 3}
Obviously, the condition (B1) is satisfied. (B2) is also satisfied as a
consequence of the fact that $||M||_{\text{tr}}\geq ||\phi (M)||_{\text{tr}}$
for any CPTP map $\phi $ and any
Hermitian matrix $M$ \cite{Petz-2006-JMP}. Concerning (B5),
we consider $\rho =p_{1}\rho _{1}\oplus p_{2}\rho _{2}$ as described in (B5).
Any $\sigma \in \mathcal{I}_{\text{BI}}(H)$ can be written as
\begin{eqnarray}
\sigma =q_{1}\sigma _{1}\oplus q_{2}\sigma _{2},  \label{eqA49}
\end{eqnarray}
with $q_{1}\geq 0$, $q_{2}\geq 0$, $q_{1}+q_{2}=1,$ and $\sigma _{1},\sigma
_{2}\in \mathcal{S}(H)$, $\sigma _{1}P_{k_{2}}=\sigma _{2}P_{k_{1}}=0$ for
any $k_{1}$ and $k_{2}.$ It follows that
\begin{eqnarray}
&&C(p_{1}\rho _{1}\oplus p_{2}\rho _{2},P) \nonumber \\
&=&\min_{\lambda >0,q_{1},\sigma _{1},\sigma _{2}}||p_{1}\rho _{1}\oplus
p_{2}\rho _{2}-\lambda (q_{1}\sigma _{1}\oplus q_{2}\sigma _{2})||_{\text{tr}%
}   \nonumber \\
&=&\min_{\lambda >0,q_{1},\sigma _{1},\sigma _{2}}(p_{1}||\rho _{1}-\frac{%
\lambda q_{1}}{p_{1}}\sigma _{1}||_{\text{tr}}+p_{2}||\rho _{2}-\frac{%
\lambda q_{2}}{p_{2}}\sigma _{2}||_{\text{tr}})  \nonumber \\
&=&p_{1}\min_{\lambda _{1}>0,\sigma _{1}}||\rho _{1}-\frac{\lambda q_{1}}{%
p_{1}}\sigma _{1}||_{\text{tr}}   \nonumber \\
&&+p_{2}\min_{\lambda _{2}>0,\sigma
_{2}}||\rho _{2}-\frac{\lambda q_{2}}{p_{2}}\sigma _{2}||_{\text{tr}}  \nonumber \\
&=&p_{1}C(\rho _{1})+p_{2}C(\rho _{2},P),  \label{eqA50} \nonumber
\end{eqnarray}
where we have used the facts that $\sigma _{1},\sigma _{2}\in
\mathcal{S}(H),$ $\{q_{1},q_{2}\}$ is a probability distribution, $\lambda
_{1}=\frac{\lambda q_{1}}{p_{1}}$ and $\lambda _{2}=\frac{\lambda q_{2}}{p_{2}}$.

\subsection{Proof of Proposition 4}
It can be proved that $C_{\text{w}}(\rho ,P)$ fulfills the conditions (B1), (B3) and (B4) by using
a similar way adopted in Ref. \cite{Bu-2018-PRA}. Here we
equivalently prove that $C_{\text{w}}(\rho ,P)$ fulfills (B1), (B2) and (B5).
(B1) is evidently satisfied. To prove (B2), suppose $\{K_{l}\}_{l}\in \mathcal{C}_{%
\text{BI}}(H)$ with $\{K_{l}\}_{l}$ a BI decomposition. Then there exists $%
\sigma \in \mathcal{I}_{\text{B}}(H)$ such that
\begin{eqnarray}
\rho &\geq& \lbrack 1-C_{\text{w}}(\rho ,P)]\sigma ,  \label{eqA51}  \nonumber \\
\sum_{l}K_{l}\rho K_{l}^{\dagger }&\geq& \lbrack 1-C_{\text{w}}(\rho
,P)]\sum_{l}K_{l}\sigma K_{l}^{\dagger }.  \label{eqA52} \nonumber
\end{eqnarray}
Since $\sum_{l}K_{l}\sigma K_{l}^{\dagger }\in \mathcal{I}_{\text{B}}(H)$, we
obtain $C_{\text{w}}(\sum_{l}K_{l}\rho K_{l}^{\dagger },P)\leq C_{\text{w}}(\rho ,P)$,
which proves that (B2) is satisfied.

To prove (B5), let us consider again $\rho =p_{1}\rho _{1}\oplus p_{2}\rho _{2}$ as described in (B5).
Then there exists $\sigma \in \mathcal{I}_{\text{B}}(H)$ such that
\begin{eqnarray}
\rho &\geq& \lbrack 1-C_{\text{w}}(\rho ,P)]\sigma ,   \label{eqA53}  \nonumber \\
\sum_{k_{1}}P_{k_{1}}\rho P_{k_{1}}&\geq& \lbrack 1-C_{\text{w}}(\rho
,P)]\sum_{k_{1}}P_{k_{1}}\sigma P_{k_{1}},    \label{eqA54}  \nonumber \\
\sum_{k_{2}}P_{k_{2}}\rho P_{k_{2}}&\geq& \lbrack 1-C_{\text{w}}(\rho
,P)]\sum_{k_{2}}P_{k_{2}}\sigma P_{k_{2}}.  \label{eqA55} \nonumber
\end{eqnarray}
Denote $\sum_{k_{1}}P_{k_{1}}\sigma P_{k_{1}}=q_{1}\sigma _{1}$, $%
\sum_{k_{1}}P_{k_{1}}\sigma P_{k_{1}}=q_{2}\sigma _{2}$, with $%
\{q_{1},q_{2}\}$ a probability distribution, $\sigma _{1},\,\sigma _{2}\in
\mathcal{I}_{\text{B}}(H)$.
Since $\sum_{k_{1}}P_{k_{1}}\rho P_{k_{1}}=p_{1}\rho _{1}$, $%
\sum_{k_{2}}P_{k_{2}}\rho P_{k_{2}}=p_{2}\rho _{2}$, we have
\begin{eqnarray}
\rho _{1}&\geq& \frac{\lbrack 1-C_{\text{w}}(\rho ,P)]q_{1}}{p_{1}}\sigma _{1},   \label{eqA56}  \nonumber \\
\rho _{2}&\geq& \frac{\lbrack 1-C_{\text{w}}(\rho ,P)]q_{2}}{p_{2}}\sigma _{2},  \label{eqA57}   \nonumber \\
C_{\text{w}}(\rho _{1},P)&\leq& 1-\frac{[1-C_{\text{w}}(\rho ,P)]q_{1}}{p_{1}}, \label{eqA58}   \nonumber \\
C_{\text{w}}(\rho _{2},P)&\leq& 1-\frac{[1-C_{\text{w}}(\rho ,P)]q_{2}}{p_{2}},  \label{eqA59}  \nonumber \\
p_{1}C_{\text{w}}(\rho _{1},P)&+&p_{2}C_{\text{w}}(\rho _{2},P)\leq C_{\text{w}}(\rho ,P). \label{eqA60}
\end{eqnarray}
Conversely, there exist $\sigma _{1}^{\prime },\sigma _{2}^{\prime }\in
\mathcal{I}_{\text{B}}(H)$ such that
\begin{eqnarray}
\rho _{1}\geq \lbrack 1-C_{\text{w}}(\rho _{1},P)]\sigma _{1}^{\prime },  \label{eqA61}   \nonumber \\
\rho _{2}\geq \lbrack 1-C_{\text{w}}(\rho _{2},P)]\sigma _{2}^{\prime }.  \label{eqA62} \nonumber
\end{eqnarray}
It follows that
\begin{eqnarray}
&&p_{1}\rho _{1}\oplus p_{2}\rho _{2}  \nonumber \\
&\geq& p_{1}[1-C_{\text{w}}(\rho _{1},P)]\sigma
_{1}^{\prime }+p_{2}[1-C_{\text{w}}(\rho _{2},P)]\sigma _{2}^{\prime },  \ \ \ \   \label{eqA63}   \nonumber \\
&&C_{\text{w}}(\rho ,P)\leq p_{1}C_{\text{w}}(\rho _{1},P)+p_{2}C_{\text{w}}(\rho _{2},P).  \label{eqA64}
\end{eqnarray}
Eqs. (\ref{eqA60}) and (\ref{eqA64}) imply (B5), which completes the proof.

\subsection{Proof of Proposition 5}
This proof is a generalization of the proof for the Theorem 1 in
Ref. \cite{Xu-2020-CPB}. For $\alpha \in \lbrack
\frac{1}{2},1),$ $\sigma,\rho \in \mathcal{S}(H),$ the sandwiched R\'{e}nyi
relative entropy is defined as \cite{Yang-2014-CMP,Muller-Lennert-2013-JMP},
\begin{eqnarray}
F_{\alpha }(\sigma ||\rho )=\frac{\ln \text{tr}[(\rho ^{\frac{1-\alpha }{%
2\alpha }}\sigma \rho ^{\frac{1-\alpha }{2\alpha }})^{\alpha }]}{\alpha -1}. \label{eqA65} \nonumber
\end{eqnarray}

It is shown that \cite{Muller-Lennert-2013-JMP,Beigi-2013-JMP} for $\alpha
\in \lbrack \frac{1}{2},1),$
$F_{\alpha }(\sigma ||\rho )\geq 0$, where the equality holds if and only if $\sigma =\rho$.
This is equivalent to that
\begin{eqnarray}
\text{tr}[(\rho ^{\frac{1-\alpha }{2\alpha }}\sigma \rho ^{\frac{1-\alpha }{%
2\alpha }})^{\alpha }]\leq 1, \label{eqA67} \nonumber
\end{eqnarray}
and to that
\begin{eqnarray}
\{\text{tr}[(\rho ^{\frac{1-\alpha }{2\alpha }}\sigma \rho ^{\frac{1-\alpha
}{2\alpha }})^{\alpha }]\}^{\frac{1}{1-\alpha }}\leq 1,\label{eqA68} \nonumber
\end{eqnarray}
with the equality holding if and only if $\sigma =\rho$.
This says that $C_{R,\alpha }(\rho ,P)$ satisfies (B1).

For $\alpha \in \lbrack \frac{1}{2},1),$ it has been shown that \cite{Muller-Lennert-2013-JMP,Lieb-2013-JMP} for $\sigma ,\rho \in \mathcal{S}(H),$
and any CPTP map $\phi ,$
\begin{eqnarray}
F_{\alpha }(\phi (\sigma )||\phi (\rho ))\leq F_{\alpha }(\sigma ||\rho ). \label{eqA69} \nonumber
\end{eqnarray}
This implies
\begin{eqnarray}
&&\text{tr}[(\phi (\rho ))^{\frac{1-\alpha }{2\alpha }}\phi (\sigma )(\phi
(\rho ))^{\frac{1-\alpha }{2\alpha }})^{\alpha }]  \nonumber \\
 &\geq& \text{tr}[(\rho ^{%
\frac{1-\alpha }{2\alpha }}\sigma \rho ^{\frac{1-\alpha }{2\alpha }%
})^{\alpha }],   \label{70}  \nonumber \\
&&\{\text{tr}[(\phi (\rho ))^{\frac{1-\alpha }{2\alpha }}\phi (\sigma )(\phi
(\rho ))^{\frac{1-\alpha }{2\alpha }})^{\alpha }]\}^{\frac{1}{1-\alpha }} \nonumber \\
&\geq& \{\text{tr}[(\rho ^{\frac{1-\alpha }{2\alpha }}\sigma \rho ^{\frac{%
1-\alpha }{2\alpha }})^{\alpha }]\}^{\frac{1}{1-\alpha }}. \label{eqA71} \nonumber
\end{eqnarray}%
For any BI map $\phi _{\text{BI}},$ there exists $\sigma ^{\ast }\in
\mathcal{I}_{\text{B}}(H)$ such that
\begin{eqnarray}
&&\ \ \max_{\sigma \in \mathcal{I}_{\text{B}}(H)}\{\text{tr}[(\rho ^{\frac{%
1-\alpha }{2\alpha }}\sigma \rho ^{\frac{1-\alpha }{2\alpha }})^{\alpha
}]\}^{\frac{1}{1-\alpha }}  \nonumber \\
&=&\{\text{tr}[(\rho ^{\frac{1-\alpha }{2\alpha }}\sigma ^{\ast }\rho ^{%
\frac{1-\alpha }{2\alpha }})^{\alpha }]\}^{\frac{1}{1-\alpha }}  \nonumber \\
&\leq &\{\text{tr}[(\phi _{\text{BI}}(\rho ))^{\frac{1-\alpha }{2\alpha }}\phi _{%
\text{BI}}(\sigma ^{\ast })(\phi _{\text{BI}}(\rho ))^{\frac{1-\alpha }{%
2\alpha }})^{\alpha }]\}^{\frac{1}{1-\alpha }}  \nonumber \\
&\leq &\max_{\sigma \in \mathcal{I}_{\text{B}}(H)}\{\text{tr}[(\phi _{\text{BI}%
}(\rho ))^{\frac{1-\alpha }{2\alpha }}\sigma (\phi _{\text{BI}}(\rho ))^{%
\frac{1-\alpha }{2\alpha }})^{\alpha }]\}^{\frac{1}{1-\alpha }}. \ \ \ \ \ \  \label{eqA72} \nonumber
\end{eqnarray}%
This proves that $C_{R,\alpha }(\rho ,P)$ satisfies (B2).

Next we prove $C_{R,\alpha }(\rho ,P)$ satisfies (B5). Consider $\rho
=p_{1}\rho _{1}\oplus p_{2}\rho _{2}$  as described in (B5). As any $\sigma \in
\mathcal{I}_{\text{BI}}(H)$ can be written as Eq. \eqref{eqA49}, it follows that
\begin{eqnarray}
&&\max_{\sigma \in \mathcal{I}_{\text{B}}(H)}\text{tr}[(\rho ^{\frac{%
1-\alpha }{2\alpha }}\sigma \rho ^{\frac{1-\alpha }{2\alpha }})^{\alpha }]\
\ \ \ \ \ \ \ \ \ \ \ \ \ \ \ \ \ \ \ \ \ \ \   \nonumber \\
&=&\max_{q_{1},q_{2}}\{(p_{1}^{1-\alpha }q_{1}^{\alpha })\max_{\sigma _{1}}%
\text{tr}[(\rho _{1}^{\frac{1-\alpha }{2\alpha }}\sigma _{1}\rho _{1}^{\frac{%
1-\alpha }{2\alpha }})^{\alpha }]\ \ \   \nonumber \\
&&+(p_{2}^{1-\alpha }q_{2}^{\alpha })\max_{\sigma _{2}}\text{tr}[(\rho _{2}^{%
\frac{1-\alpha }{2\alpha }}\sigma _{2}\rho _{2}^{\frac{1-\alpha }{2\alpha }%
})^{\alpha }]\}  \nonumber \\
&=&\max_{q_{1},q_{2}}\{p_{1}^{1-\alpha }q_{1}^{\alpha }t_{1}+p_{2}^{1-\alpha
}q_{2}^{\alpha }t_{2}\}\ \ \ \ \ \ \ \ \ \ \ \ \ \ \ \   \nonumber \\
&=&p_{1}^{1-\alpha }p_{2}^{1-\alpha }t_{1}t_{2}(p_{1}^{-1}t_{1}^{\frac{1}{%
\alpha -1}}+p_{2}^{-1}t_{2}^{\frac{1}{\alpha -1}})^{1-\alpha },\ \  \label{eqA73} \nonumber
\end{eqnarray}%
where
\begin{eqnarray}
t_{1} &=&\max_{\sigma _{1}}\text{tr}[(\rho _{1}^{\frac{1-\alpha }{2\alpha }%
}\sigma _{1}\rho _{1}^{\frac{1-\alpha }{2\alpha }})^{\alpha },  \label{eqA74}  \nonumber \\
t_{2} &=&\max_{\sigma _{2}}\text{tr}[(\rho _{2}^{\frac{1-\alpha }{2\alpha }%
}\sigma _{2}\rho _{2}^{\frac{1-\alpha }{2\alpha }})^{\alpha }],  \label{eqA75} \nonumber
\end{eqnarray}%
and the Lemma 1 (note here $t_{1}>0$ and $t_{2}>0)$ has been taken into account.

Consequently,
\begin{eqnarray}
&&\max_{\sigma \in \mathcal{I}_{\text{B}}(H)}(\{\text{tr}[(\rho ^{\frac{1-\alpha }{%
2\alpha }}\sigma \rho ^{\frac{1-\alpha }{2\alpha }})^{\alpha }]\}^{\frac{1}{%
1-\alpha }})\ \ \ \ \ \ \ \ \ \   \nonumber \\
&=&\{\max_{\sigma \in \mathcal{I}_{\text{B}}(H)}\text{tr}[(\rho ^{\frac{1-\alpha }{%
2\alpha }}\sigma \rho ^{\frac{1-\alpha }{2\alpha }})^{\alpha }]\}^{^{\frac{1%
}{1-\alpha }}}\ \ \ \ \ \ \ \ \ \ \   \nonumber \\
&=&p_{1}p_{2}t_{1}^{\frac{1}{1-\alpha }}t_{2}^{\frac{1}{1-\alpha }%
}(p_{1}^{-1}t_{1}^{\frac{1}{\alpha -1}}+p_{2}^{-1}t_{2}^{\frac{1}{\alpha -1}%
})\ \ \   \nonumber \\
&=&p_{1}t_{1}^{\frac{1}{1-\alpha }}+p_{2}t_{2}^{\frac{1}{1-\alpha }}.\ \ \ \
\ \ \ \ \ \ \ \ \ \ \ \ \ \ \ \ \ \ \ \ \ \  \label{eqA76} \nonumber
\end{eqnarray}%
This shows that $C_{R,\alpha }(\rho ,P)$ satisfies (B5).

%\bibliographystyle{apsrev4-1}
%\bibliography{POVMcoherence}
%

\end{document}